\documentclass[amsmath,amssymb,prc,final,showpacs,twocolumn]{revtex4-1}
\usepackage{bm,hyphenat,xspace}
\usepackage{graphicx,epsfig,amssymb}

\newcommand {\mcu}{\mathcal{U}}
\newcommand {\mct}{\mathcal{T}}

\newcommand{\cm}{\mathrm{c\!\:\!.m\!\:\!.}}
\newcommand{\He}{{}^3\mathrm{He}}
\newcommand{\Hh}{{}^3\mathrm{H}}
\newcommand{\Hd}{{}^2\mathrm{H}}

\begin{document}

\title 
{Proton-${}^3$H scattering calculation: \\ Elastic and charge-exchange reactions
up to 30 MeV}
 
\author{A.~Deltuva} 
\affiliation
{Institute of Theoretical Physics and Astronomy, 
Vilnius University, A. Go\v{s}tauto 12, LT-01108 Vilnius, Lithuania
}

\author{A.~C.~Fonseca} 
\affiliation
{Centro de F\'{\i}sica Nuclear da Universidade de Lisboa, 
P-1649-003 Lisboa, Portugal }

\received{\today}

\begin{abstract}
Proton-${}^3$H  elastic scattering and charge-exchange reaction $\Hh(p,n)\He$ 
 in the energy regime above four-nucleon breakup threshold are described 
in the momentum-space transition operator framework. Fully converged results are 
obtained using realistic two-nucleon
potentials and two-proton  Coulomb force  as dynamic input.
 Differential cross section, proton analyzing power,
outgoing neutron polarization, and proton-to-neutron polarization
transfer coefficients are calculated between 6 and 30 MeV proton beam energy.
Good agreement with the  experimental data is found for the
differential cross section both in elastic and  charge-exchange reactions; the latter
shows a complicated energy and angular dependence. The most sizable discrepancies
between predictions and data 
are found for the  proton analyzing power and outgoing neutron polarization
in the charge-exchange reaction, while the respective proton-to-neutron polarization
transfer coefficients are well described by the calculations.
\end{abstract}

 \maketitle

\section{Introduction \label{sec:intro}}

The theoretical understanding of the structure of nuclei
along the valley of stability, as well as away from it up to the
neutron or proton drip lines, has advanced fast in the last 15 years
through state-of-the-art microscopic Green's function Monte Carlo
(GFMC)  \cite{pieper:01a,pieper:02a} and No Core Shell Model (NCSM)
\cite{caurier:02a} calculations based on realistic
nucleon-nucleon ($NN$) and three-nucleon $(3N)$ force models.
 In contrast, the corresponding advances in the study of nuclear
reactions have been meager and mostly limited to the three- and four
-nucleon ($4N$) systems. As discussed in a recent review article 
\cite{carbonell:14a},
this may change in the near future as one implements algorithms
capable of applying bound state techniques to the solution of the
multiparticle scattering problem. Until this becomes a reliable
pathway, one follows the traditional approach by solving coordinate- or
momentum-space equations with appropriate boundary conditions that are
equivalent to solving the corresponding $n$-particle Schr\"odinger
equation.

Although rigorous $n$-particle scattering equations were derived 
almost 50 years ago by Faddeev and Yakubovsky (FY)  
 \cite{faddeev:60a,yakubovsky:67} and by
 Alt, Grassberger, and Sandhas (AGS) \cite{alt:67a,grassberger:67},
exact  numerical solutions of the $3N$ and $4N$
scattering problems only became possible with the advent of fast and
larger computers, together with powerful numerical techniques such as
spline interpolation, Pad\'{e} summation,
 and many others. Neutron-deuteron ($n$-$d$) scattering calculations with
realistic $NN$ force models reached state-of-the-art status in the early
1990's due to the effort of a number of independent
groups \cite{koike:86a,chen:89a,friar:89a,cornelius:90a,kievsky:96a}.
Owing to the difficulties in treating the long-range Coulomb force, fully
converged proton-deuteron  ($p$-$d$) scattering calculations came later
\cite{kievsky:01a,chen:01a,deltuva:05a,deltuva:05d}. 
 Due to its higher dimensionality and multichannel complexity, the
$4N$ scattering problem took  twenty years longer to reach
the same status as the $3N$ system except for the calculation
of breakup reactions. There have been also attempts to calculate 
 scattering processes involving five and more nucleons but using 
different methods
than in the $3N$ and $4N$ systems, namely, GFMC \cite{nollett:07a}
and the NCSM resonating group \cite{quaglioni:08a}.

Although the $4N$ system has a long history that started out in the
early 1970's \cite{fonseca:87}, most of the recent developments
are mainly due to the works of the
Pisa \cite{viviani:01a,kievsky:08a,viviani:10a,viviani:13a},
Grenoble-Strasbourg
\cite{lazauskas:phd,lazauskas:04a,lazauskas:09a,lazauskas:12a}, and
Lisbon \cite{deltuva:07a,deltuva:07b,deltuva:12c,deltuva:13c}
groups. Because the first two groups use the coordinate-space
representation, they were able to include not only realistic $NN$
interactions but also $3N$ forces. Nevertheless,
they have had a major difficulty in calculating multichannel reactions
and going beyond breakup threshold, particularly when the Coulomb
interaction between protons  is included. The Lisbon group uses the
momentum-space AGS equations for
transition operators \cite{grassberger:67} that were solved for
multichannel reactions both below \cite{deltuva:07c}
and above \cite{deltuva:14a} breakup threshold and with the
Coulomb force included. The only stumbling block has been the
inclusion of irreducible $3N$ forces.
For this reason we are not yet able
to perform calculations with $NN$ and $3N$ 
potentials derived from the chiral effective field theory;
 only the $NN$ part has been included in our calculations
\cite{deltuva:07a,deltuva:07c}.
An alternative is a
nuclear force model with explicit excitation of a nucleon to a
$\Delta$ isobar. This coupling generates both effective $3N$ and
$4N$ forces that have been successfully included
in $4N$ calculations by the Lisbon-Hannover collaboration
\cite{deltuva:08a}.

In the past two years we  produced
a realm of results for cross sections and polarizations observables
for the elastic scattering of a
neutron ($n$) on a $\Hh$ target \cite {deltuva:12c}
 and a proton ($p$) on a $\He$ target \cite{deltuva:13c}
above the breakup threshold. These processes are dominated by 
the total isospin $\mct=1$ states.  More recently we
presented results for the mixed isospin  ($\mct=0$ and 1) processes
initiated by the  $n+\He$ collisions \cite{deltuva:14b}
that are coupled multichannel reactions leading to all energetically
allowed final states  $n+\He$, $p+\Hh$, $d+d$, $d+n+p$, and $n+n+p+p$.

In the present work we study in detail the $p+\Hh$ scattering
above the breakup threshold. We concentrate on  
elastic and charge-exchange reactions for which there is abundant 
experimental data, but also some  inconsistencies
between different data sets, that might be sorted out through accurate
theoretical predictions. 
These calculations are especially important given that novel experiments 
are hardly possible due to the lack of $\Hh$ targets or proper
facilities that still operate at these energies. Furthermore, new $4N$
scattering calculations are also worth pursuing because they lead the way
to the solution of complicated multiparticle scattering problems, not
just in nuclear physics but also in cold atom physics
\cite{deltuva:10c}, and in the study of complex nuclear reactions that exhibit
four-body degrees of freedom such as the scattering of a two-neutron
halo nucleus on a proton target.
 
In Sec.~\ref{sec:eq} we shortly recall the theoretical formalism
and in Sec.~\ref{sec:res} we present the numerical results.
The summary is given in  Sec.~\ref{sec:sum}.

\section{4N scattering equations \label{sec:eq}}

We employ the isospin formalism for the description of the $4N$ scattering.
Since  $p+\Hh$ is the mirror of the $n+\He$ system, we take over 
the calculational technique from Ref.~\cite{deltuva:14b}
where $4N$ reactions initiated by the $n+\He$ collisions were described.
Thus, we use the momentum-space partial-wave framework to
solve the  integral AGS equations \cite{grassberger:67} 
 for symmetrized four-particle transition operators 
\begin{subequations}  \label{eq:AGS}   
\begin{align}  
\mcu_{11}  = {}&  -(G_0 \, t \, G_0)^{-1}  P_{34} -
P_{34} U_1 G_0 \, t \, G_0 \, \mcu_{11}  \nonumber \\ {}& 
+ U_2   G_0 \, t \, G_0 \, \mcu_{21}, \label{eq:U11}  \\
\label{eq:U21}
\mcu_{21} = {}&  (G_0 \, t \, G_0)^{-1}  (1 - P_{34})
+ (1 - P_{34}) U_1 G_0 \, t \, G_0 \, \mcu_{11}.
\end{align}
\end{subequations}
Here $G_0= (E+ i\varepsilon - H_0)^{-1}$ is the free resolvent
with the complex energy $E+ i\varepsilon$ and the free Hamiltonian $H_0$,
$t$ is the $NN$ transition matrix,
$P_{ab}$ is the  permutation operator of particles $a$ and $b$,
and  $U_\alpha$ are the transition operators for the 3+1 ($\alpha=1$)
and 2+2 ($\alpha=2$) subsystems. 
The on-shell matrix elements of the operators $\mcu_{\beta\alpha} $
taken at  $\varepsilon \to +0$ yield the transition amplitudes for two-cluster reactions.
In the isospin formalism the $NN$ transition matrix has contributions from
$nn$, $np$, and $pp$ pairs with the respective weights given in 
 Ref.~\cite{deltuva:14b}; for the $pp$ pair the screened Coulomb interaction
is included and the resulting physical amplitudes for $4N$ reactions are 
obtained by using the method of screening and renormalization
\cite{taylor:74a,alt:80a,deltuva:05a,deltuva:07c}.
Further explanations and technical details can be found in 
Refs.~\cite{deltuva:14a,deltuva:14b}.

\begin{figure}[!]
\begin{center}
\includegraphics[scale=0.69]{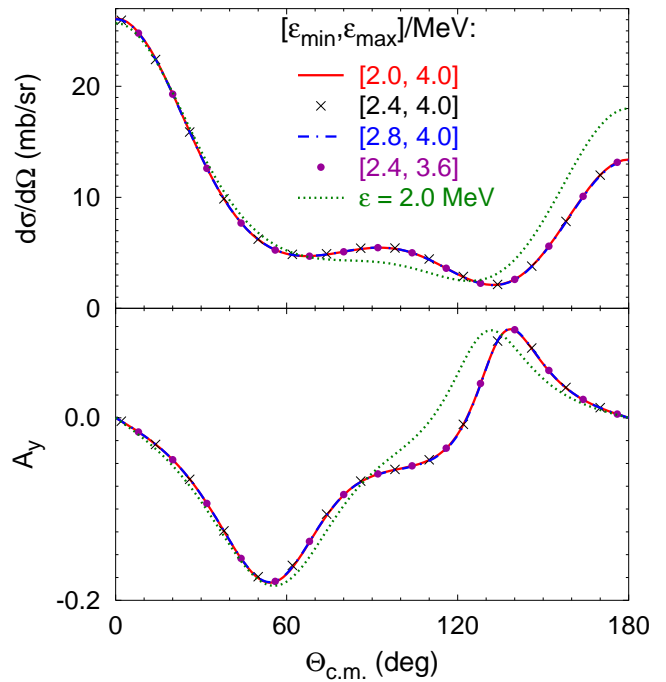}
\end{center} 
\caption{ \label{fig:conv} (Color online)
Differential cross section and proton analyzing power for
$\Hh(p,n)\He$ reaction at 24 MeV proton energy
as functions of the c.m. scattering angle.
Results obtained using different sets of $\varepsilon$ 
values ranging from $\varepsilon_{\mathrm{min}}$ to 
$\varepsilon_{\mathrm{max}}$ 
with the step of 0.4 MeV are compared; they are indistinguishable.
The dotted curves refer to the  $\varepsilon = 2.0$ MeV calculations
without extrapolation that have no physical meaning but
show the importance of the extrapolation. }
\end{figure}

To avoid the very complicated singularity structure of the kernel, we solve
Eqs.~\eqref{eq:AGS}  numerically at several complex energies $E+ i\varepsilon_j$
with  finite values of $\varepsilon_j > 0$ and then 
extrapolate the obtained on-shell matrix elements of the 
 transition operators $\mcu_{\beta\alpha} $ to the $\varepsilon \to +0$ limit
which corresponds to  the physical scattering process;
more details are given in  Ref.~\cite{deltuva:12c}
using  $n+\Hh$ elastic scattering as an example.
In Fig.~\ref{fig:conv} we demonstrate that the employed method is accurate and reliable also 
for inelastic reactions. As an example we consider the charge-exchange reaction  $\Hh(p,n)\He$
at 24 MeV proton energy. The transition amplitudes were calculated at six values
of $\varepsilon_j > 0$ ranging from 2 to 4 MeV, and then extrapolated to the
 $\varepsilon \to +0$ limit using four different sets. In all four cases the resulting
differential cross section and proton analyzing power turn out to be indistinguishable in the 
plot, whereas the predictions at finite $\varepsilon=2$ MeV  without extrapolation 
deviate significantly.

\section{Results \label{sec:res}}

\begin{figure*}[!]
\begin{center}
\includegraphics[scale=0.7]{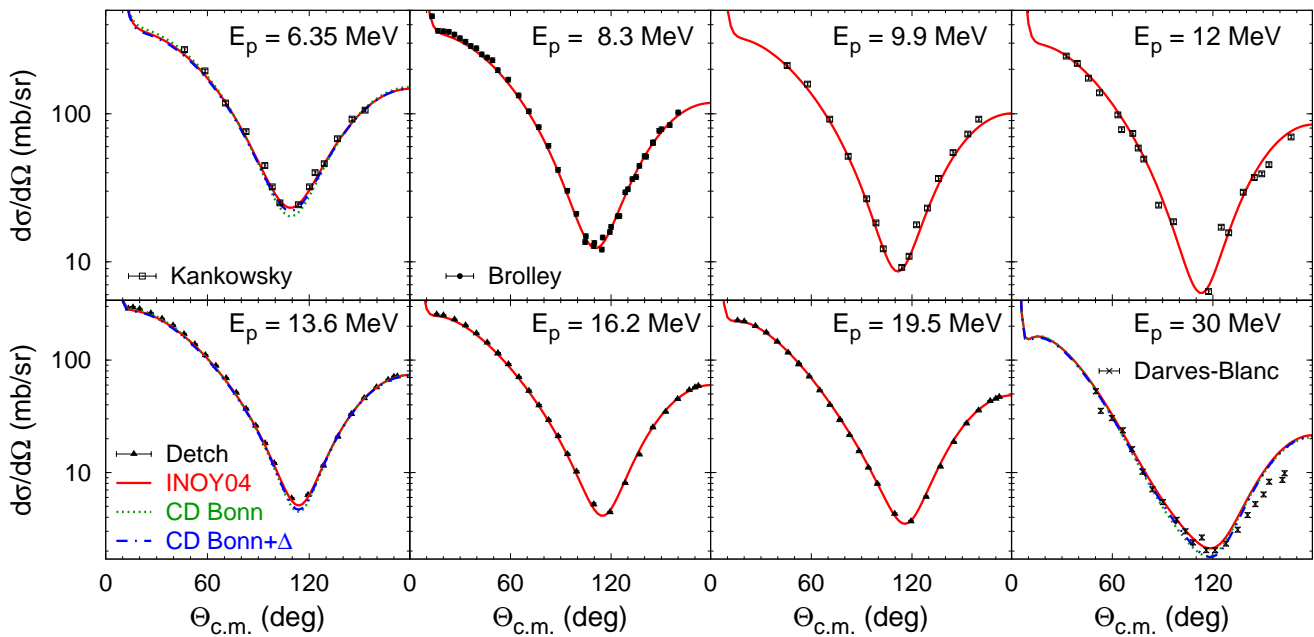}
\end{center}
\caption{\label{fig:dcs} (Color online) Differential cross section 
 of elastic $p+\Hh$ scattering at proton energy between 6.35 and 30 MeV.
  Results obtained with potentials INOY04
  (solid curves), CD Bonn (dotted curves), and CD Bonn + $\Delta$ (dashed-dotted curves) are
  compared with data  from 
Refs.~\cite{kankowsky:76,brolley:60,PhysRevC.4.52,pt19-57}.}
\end{figure*}

We consider $p+\Hh$ scattering at proton energies $E_p$ ranging from 6 to 30 MeV;
the regime below 6 MeV was studied by us in Ref.~\cite{deltuva:07c}.
Most results are obtained using  the realistic inside-nonlocal outside-Yukawa
(INOY04) potential  by Doleschall \cite{doleschall:04a,lazauskas:04a}. It
 predicts  the $\He$ ($\Hh$) binding energy to be 
7.73 MeV (8.49 MeV) and thereby nearly reproduces the experimental value of 7.72 MeV (8.48 MeV)
without an additional $3N$ force. To investigate the dependence of our results on the interaction model,
at several energies we also show the predictions obtained with
other high-precision $NN$ potentials, namely, the 
 charge-dependent Bonn potential (CD Bonn)  \cite{machleidt:01a}
and its extension CD Bonn + $\Delta$ \cite{deltuva:03c}
explicitly including an excitation of a nucleon to a $\Delta$ isobar.
This mechanism generates  effective $3N$ and $4N$ forces that are 
mutually consistent but quantitatively still insufficient to reproduce
$3N$ and $4N$ binding energies, although they reduce the discrepancy \cite{deltuva:08a}.
For $\He$ and $\Hh$ binding energies  the  CD Bonn + $\Delta$
potential yields 7.53  and 8.28 MeV, while the predictions of CD Bonn are
 7.26 and 8.00 MeV, respectively.
In addition to INOY04,
results for CD Bonn + $\Delta$  are presented at
$E_p = 6.0$, 6.35,  13.6, and 30.0 MeV while 
for CD Bonn at $E_p = 6.0$, 6.35, 7.0, 9.0, 13.0, 13.6, 21.0, and 30.0 MeV.

\begin{figure*}[!]
\begin{center}
\includegraphics[scale=0.7]{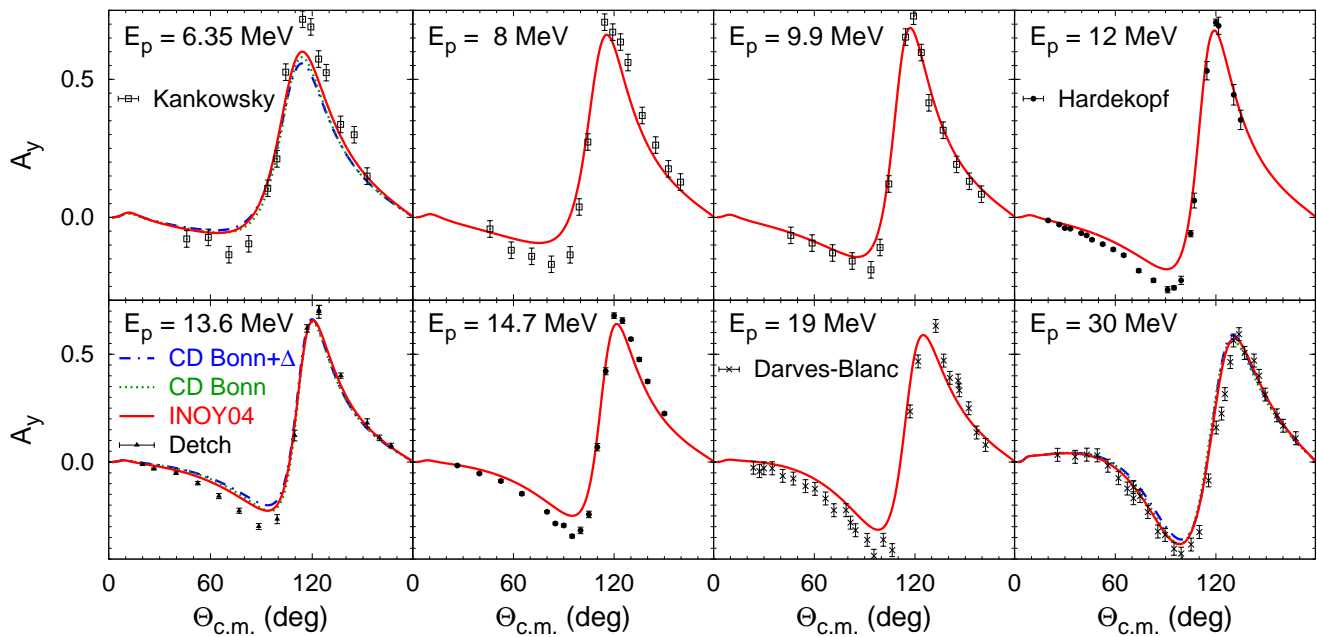}
\end{center}
\caption{\label{fig:ay} (Color online) 
  Proton analyzing power of elastic $p+\Hh$ scattering at
  proton  energy between 6.35 and 30 MeV.  
Curves are  as in Fig.~\ref{fig:dcs}. Data are  from 
Refs.~\cite{kankowsky:76,hardekopf:72a,PhysRevC.4.52,pt19-57}.}
\end{figure*}

In Fig.~\ref{fig:dcs}  we show the 
differential cross section $d\sigma/d\Omega$ for elastic $p+\Hh$ scattering
as a function of the center of mass (c.m.) scattering angle $\Theta_{\cm}$.
This observable decreases with the increasing energy 
while its minimum moves slowly to higher angles; the INOY04 calculations
describe the energy and angular dependence of the experimental data
\cite{kankowsky:76,hardekopf:72a,PhysRevC.4.52,pt19-57}
fairly well except at backward angles at 30 MeV. However, one may question the 
reliability of those data \cite{pt19-57} that exhibit quite an abrupt decrease 
from $E_p = 19.5$ MeV to  $E_p = 30$ MeV while other data and theoretical predictions
vary smoothly with energy. Unfortunately, we found no data between $E_p = 20$ and 30 MeV
that would help sort out this possible discrepancy.
The sensitivity to the force model manifests itself in the minimum where
predictions obtained with CD Bonn and CD Bonn + $\Delta$ are below those of INOY04 by about
15 \%.

It is interesting to compare the present $p+\Hh$ results with those
for $n+\Hh$ \cite{deltuva:12c},  $p+\He$ \cite{deltuva:13c}, and
$n+\He$  \cite{deltuva:14b}  elastic scattering. Since the energy and force model
dependence for $p+\Hh$ and $n+\He$ calculations are the same as can be expected
from the isospin symmetry, one
would expect similar discrepancies between data and theory assuming that
the data can be equally trusted. However,
 the agreement with the experimental data is different in these two cases.
We note that for $p+\Hh$ there
is no discrepancy at the minimum of the differential cross section as the proton
energy increases towards 30 MeV, while in $n+\He$  and $p+\He$ elastic scattering
the minimum is  underpredicted by the  theoretical
calculations as the energy of the incoming beam rises above 25 MeV. In contrast,
the $n+\Hh$ data, only available at 22 MeV, are overpredicted \cite{deltuva:12c}.

In Fig.~\ref{fig:ay}  we show the proton analyzing power  $A_{y}$
for the elastic $p+\Hh$ scattering 
at proton energies ranging from 6.35 to 30 MeV. 
The qualitative reproduction of the experimental data by our calculations
is reasonable. The existing discrepancies around the minimum and the maximum decrease 
as the energy increases reaching a good agreement at $E_p = 30$ MeV.
 The sensitivity to the nuclear force model  is  quite weak. 
In all these respects the behavior of the 
$A_{y}$ in the elastic $p+\Hh$ scattering is qualitatively the same
as observed for the neutron analyzing power in the
$n+\He$ elastic scattering  \cite{deltuva:14b}.

\begin{figure*}[!]
\begin{center}
\includegraphics[scale=0.7]{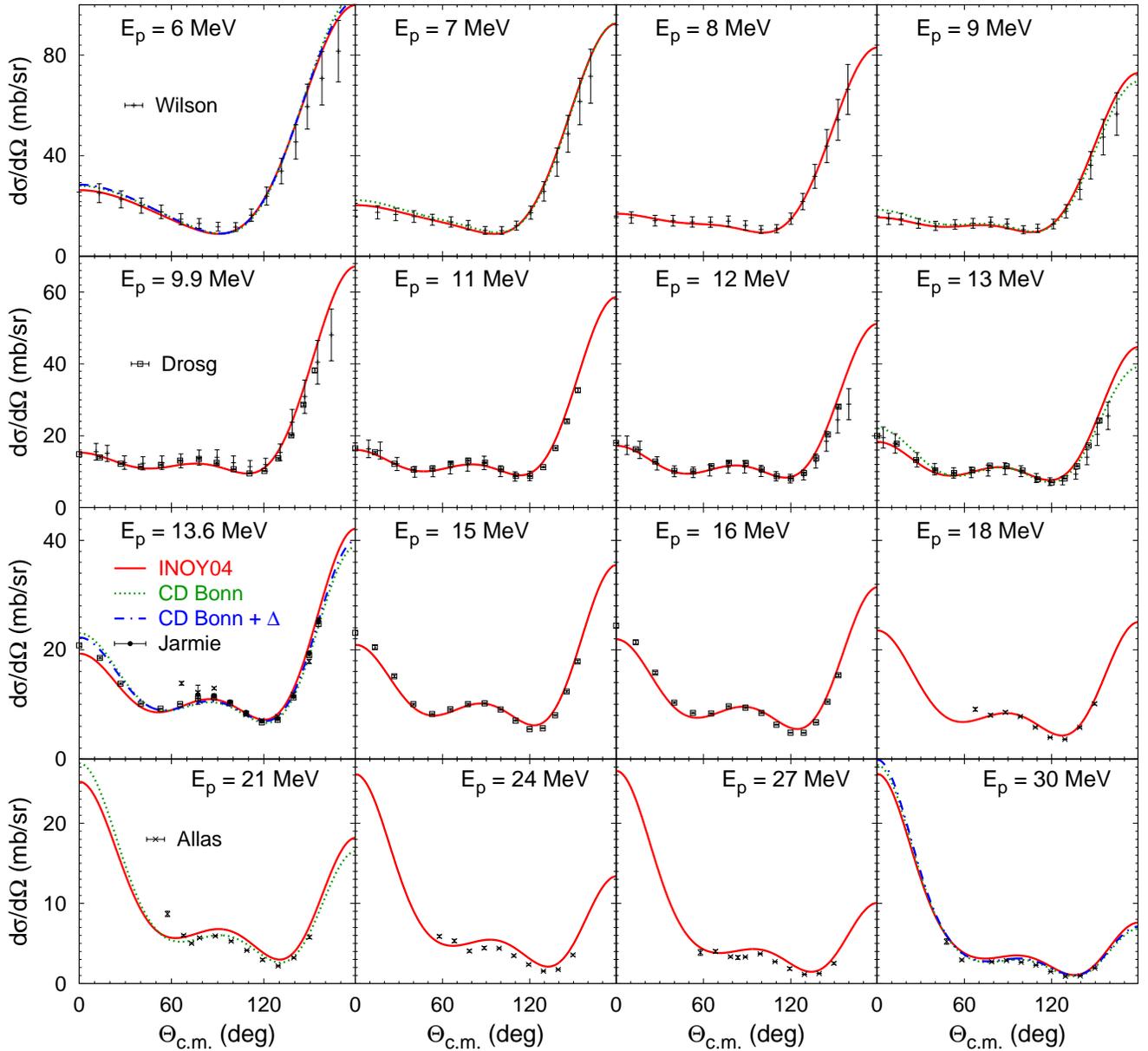}
\end{center}
\caption{\label{fig:nh-pt} (Color online) Differential cross section
  of  $\Hh(p,n)\He$ reaction. Curves are as in Fig.~\ref{fig:dcs}.
Data are from  Refs.~\cite{wilson:61,drosg:78,PhysRevC.16.15,allas:74a}. }
\end{figure*}

Next we consider rearrangement  reactions initiated by $p+\Hh$
collisions. In Fig.~\ref{fig:nh-pt} we show the differential cross
section $d\sigma/d\Omega$ for the charge exchange reaction
$\Hh(p,n)\He$ at $E_p$ ranging from 6 to 30 MeV. The data 
\cite{wilson:61,drosg:78,PhysRevC.16.15,allas:74a} exhibits a
strong energy dependency that is well reproduced by the
theoretical calculations, in particular the shape of the observable
that starts out backward peaked at 6 MeV with a single shallow minimum
around $90^\circ$, and ends up forward peaked at 30 MeV with two 
minima around $60^\circ$ and $140^\circ$. The appearance of a local
maximum and two local minima in both theory and data takes place at $E_p = 9$ MeV.
Predictions of the three employed force models
follow the same trend but in particular regimes may differ by almost 20 \%
as happens at $E_p = 13.6$ MeV and $\Theta_{\cm} = 0^\circ$.
At lower energies the sensitivity to the $NN$ potential shows up at
forward and backward angles  while above 20 MeV it extends to the whole angular regime,
being around 10 \%. The effect seems to be more complicated than just  a simple scaling
with $3N$ binding energy since the predictions of CD Bonn and 
CD Bonn + $\Delta$ stay quite close together but deviate more from INOY04. 
While at lower energies the INOY04 potential is favored by the data,
above 20 MeV the 
 best description is provided by the CD Bonn potential whereas INOY04 overpredicts the data.
In addition, there are also some inconsistencies between different data sets,
e.g., the first few points from Ref.~\cite{allas:74a} seem to be wrong when compared to
other measurements \cite{wilson:61,drosg:78,PhysRevC.16.15}.

\begin{figure*}[!]
\begin{center}
\includegraphics[scale=0.75]{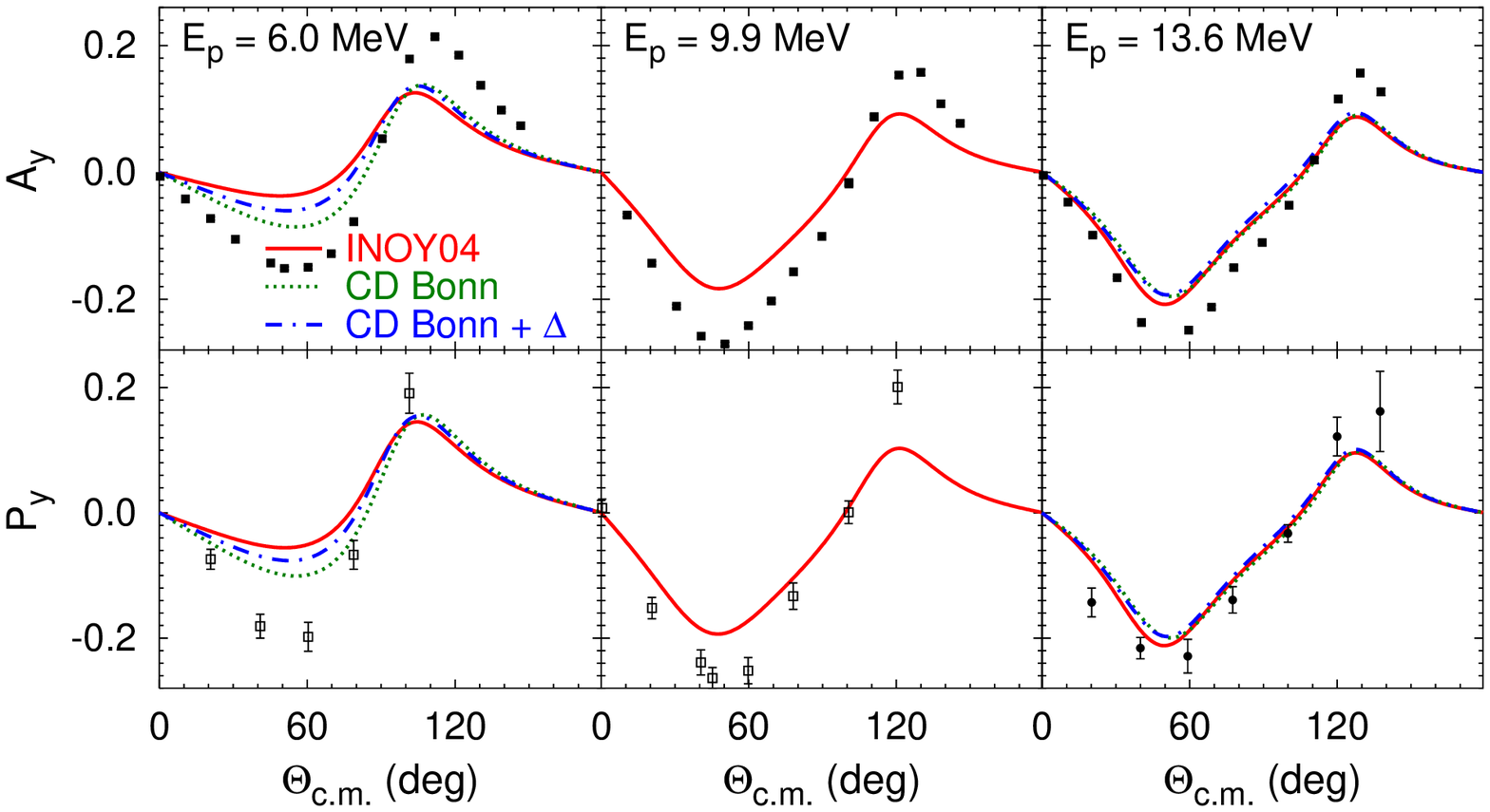}
\end{center}
\caption{\label{fig:pt-nhap} (Color online) Proton analyzing power $A_y$ and
  outgoing neutron polarization  $P_y$ in the $\Hh(p,n)\He$ reaction at 6.0,
  9.9, and 13.6 MeV proton energy.  Curves are as in Fig.~\ref{fig:dcs}.
  The  data are from Ref.~\cite{jarmer:74} for $A_y$ and
from Refs.~\cite{PhysRevC.10.494} ($\Box$) and \cite{PhysRevC.5.1826} ($\bullet$)
for  $P_y$.}
\end{figure*}

Unlike the differential cross section, the proton analyzing power $A_y$ and
the outgoing neutron polarization  $P_y$   for the 
charge-exchange reaction $\Hh(p,n)\He$ 
displayed in Fig.~\ref{fig:pt-nhap} show a large quantitative disagreement
between theoretical results and experimental data, especially at the lowest
considered energy $E_p = 6$ MeV. Here the predicted shape of these observables is 
roughly correct but the absolute value is too small by a factor of 2. 
 The disagreement decreases with increasing energy but for $A_y$ still remains about 25 \% 
at   $E_p = 13.6$ MeV, the highest energy where data are available.
Thus, nucleon vector polarization observables    $A_y$ and $P_y$ 
in the charge-exchange reaction $\Hh(p,n)\He$ exhibit 
one of the largest discrepancies seen so far in the $4N$ system.
The sensitivity to the force model is significant only at the lowest energy
where, in contrary to what can be expected from scaling with $3N$ binding energy,
the discrepancy is largest for INOY04 and smallest for CD Bonn.
This sensitivity as well as the very strong energy dependence of  $A_y$ 
observed in Ref.~\cite{deltuva:07c} might be due  to the interplay of
 $P$-wave $4N$ resonant states existing at low energies. Away from this resonant regime
the observables become less sensitive, and 
at $E_p = 13.6$ MeV the predictions of INOY04 are slightly closer to the data than those
of other potentials.
The $A_y$ and $P_y$ discrepancies observed in  the $\Hh(p,n)\He$ reaction
may be one more manisfestation of the famous $A_y$ puzzle 
seen in the elastic nucleon-deuteron and nucleon-trinucleon scattering
\cite{cornelius:90a,kievsky:01a,viviani:01a,viviani:13a,deltuva:07b,deltuva:13c}.
However, there is also an important difference: the correlation between
the discrepancy and the predicted $3N$ binding energy 
in the charge-exchange reaction below 10 MeV gets reversed 
as compared to all elastic processes. For example,
the INOY04 potential shows the smallest discrepancy in the elastic scattering
but the largest one in the  $\Hh(p,n)\He$ reaction.

\begin{figure*}[!]
\begin{center}
\includegraphics[scale=0.75]{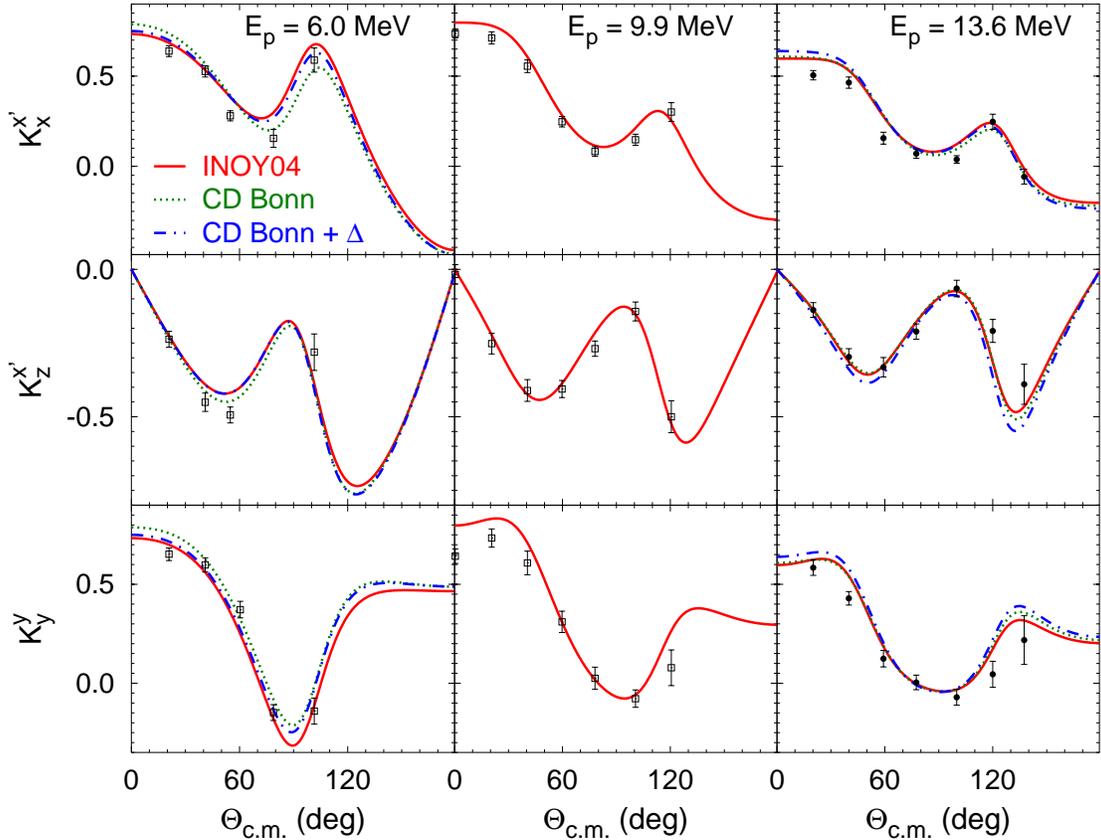}
\end{center}
\caption{\label{fig:pt-nhk} (Color online) Proton-to-neutron
  polarization transfer  coefficients of $\Hh(p,n)\He$ reaction at
  6.0, 9.9, and 13.6 MeV proton energy.   Curves are as in
  Fig.~\ref{fig:dcs}.   The data are from
   Refs.~\cite{PhysRevC.10.494} ($\Box$) and \cite{PhysRevC.5.1826} ($\bullet$).}
\end{figure*}

Although "two wrongs do not necessarily make one right", 
the proton-to-neutron polarization (spin) transfer coefficients 
$K_x^{x'}$, $K_z^{x'}$, and $K_y^{y}$ for the
charge-exchange reaction $\Hh(p,n)\He$ at $E_p = 6.0$, 9.9, and 13.6 MeV presented in
Fig.~\ref{fig:pt-nhk} are well reproduced by the theoretical calculations.
The spin transfer coefficients
  show rather complicated angular and energy dependence 
which is nonmonotonic as can be seen most clearly 
in $K_x^{x'}$ and $K_y^{y}$ at small angles. The sensitivity
of these  observables to the choice of the $NN$ force model 
is moderate over the whole considered energy and angular regime.
It is not unusual in few-nucleon physics that double-polarization observables such as 
spin correlation or spin transfer coefficients are in better agreement
with the experimental data than analyzing powers;  $p+\He$ elastic scattering 
\cite{deltuva:07b,deltuva:13c} is a further example.
Unfortunately, double-polarization data are missing in
 $n+\He$ and  $p+\Hh$ elastic scattering.

Regarding the transfer reaction $\Hh(p,d)\Hd$, first results for the 
differential cross section at $E_p = 13.6$ MeV were presented in
Ref.~\cite{deltuva:14a}; quite a good agreement between predictions and data was found with
only slight underestimation around  $\Theta_{\cm} = 90^\circ$.
Since the experimental studies  of nucleon transfer processes in the $4N$ system 
 have been dominated by the time reversed reaction $\Hd(d,p)\Hh$ and
its mirror partner $\Hd(d,n)\He$, their theoretical analysis will be presented 
in the forthcoming work on the $d+d$ scattering \cite{deltuva:15a}.

\section{Summary \label{sec:sum}}

In this work we studied $p+\Hh$ elastic scattering and
the charge-exchange reaction $\Hh(p,n)\He$  up to 30 MeV beam energy. 
We solved, in a numerically
converged way, the momentum-space Alt, Grassberger, and Sandhas
equations for transition operators.  The employed complex-energy method with special
integration weights proved to be highly reliable also for inelastic
reactions such as $\Hh(p,n)\He$.
The calculations include the Coulomb
interaction between protons together with realistic $NN$ force models, i.e.,
INOY04, CD Bonn,  and  CD Bonn + $\Delta$. An explicit excitation of 
a nucleon to a $\Delta$ isobar included in the latter model yields
mutually consistent effective $3N$  and $4N$ forces. Moderate sensitivity to
the employed interaction model is found in several observables
in particular energy regimes, most of them referring to
the $\Hh(p,n)\He$ reaction. 
Like the previous work on  $n+\Hh$,  $p+\He$, and  $n+\He$ scattering
\cite{deltuva:12c,deltuva:13c,deltuva:14a,deltuva:14b},
this is a state-of-the-art calculation
that shows the virtues and limitations of realistic $NN$ force models in
describing the world data up to 30 MeV beam energy for elastic and
charge-exchange reactions initiated by  $p+\Hh$ collisions.  We
find that elastic and charge-exchange differential cross sections are
well described by the calculations in the considered energy regime
between 6 and 30 MeV. The absence of a discrepancy in the minimum of the
elastic differential cross section at 30 MeV is quite surprising
given that such discrepancies show up in $p+\He$ and  $n+\He$ elastic scattering.
In the $\Hh(p,n)\He$ case the predicted differential cross section 
varies very rapidly with the energy, developing new local minima and maxima
in the angular distribution that are seen also in the experimental data.
 The elastic proton analyzing power $A_y$ is
fairly well described by the calculations showing the usual
discrepancies in the minima and maxima already observed in other 
elastic $4N$ collisions driven by proton or neutron beams.
The largest discrepancies between data and
calculations are observed in the proton analyzing power $A_y$ and  outgoing
neutron polarization $P_y$ in the charge-exchange reaction
$\Hh(p,n)\He$.  In contrast, 
proton-to-neutron polarization transfer coefficients in the charge-exchange reaction 
are successfully described by theory in spite of
their complex structure and variation with beam energy.  This and
previous achievements  show that, after many years of hard work, numerically
converged solutions of the $4N$ scattering problem with
realistic $NN$ force models are not only possible but also that such
endeavor has reached a level of sophistication and reliability only
comparable to $3N$ scattering studies. Nevertheless, much
remains to be done, such as calculating breakup observables or
including irreducible $3N$ forces in momentum-space calculations. Progress in
this direction is forthcoming.


\begin{thebibliography}{10}

\bibitem{pieper:01a}
S.~C. Pieper, V.~R. Pandharipande, R.~B. Wiringa, and J. Carlson, Phys.~Rev.~C
  {\bf 64},  014001  (2001).

\bibitem{pieper:02a}
S.~C. Pieper, K. Varga, and R.~B. Wiringa, Phys. Rev. C {\bf 66},  044310
  (2002).

\bibitem{caurier:02a}
E. Caurier, P. Navr\'atil, W.~E. Ormand, and J.~P. Vary, Phys. Rev. C {\bf 66},
   024314  (2002).

\bibitem{carbonell:14a}
J. Carbonell, A. Deltuva, A. Fonseca, and R. Lazauskas, Progress in Particle
  and Nuclear Physics {\bf 74},  55   (2014).

\bibitem{faddeev:60a}
L.~D. Faddeev, Zh.~Eksp.~Teor.~Fiz. {\bf 39},  1459  (1960) [Sov.~Phys. JETP
  {\bf 12}, 1014 (1961)].

\bibitem{yakubovsky:67}
O.~A. Yakubovsky, Yad. Fiz. {\bf 5},  1312  (1967) [Sov. J. Nucl. Phys. {\bf
  5}, 937 (1967)].

\bibitem{alt:67a}
E.~O. Alt, P. Grassberger, and W. Sandhas, Nucl.~Phys. {\bf B2},  167  (1967).

\bibitem{grassberger:67}
P. Grassberger and W. Sandhas, Nucl. Phys. {\bf B2},  181  (1967); E. O. Alt,
  P. Grassberger, and W. Sandhas, JINR report No. E4-6688 (1972).

\bibitem{koike:86a}
Y. Koike and Y. Taniguchi, Few-Body Systems {\bf 1},  13  (1986).

\bibitem{chen:89a}
C.~R. Chen, G.~L. Payne, J.~L. Friar, and B.~F. Gibson, Phys. Rev. C {\bf 39},
  1261  (1989).

\bibitem{friar:89a}
J.~L. Friar {\it et~al.}, Phys. Rev. C {\bf 42},  1838  (1990).

\bibitem{cornelius:90a}
T. Cornelius, W. Gl\"ockle, J. Haidenbauer, Y. Koike, W. Plessas, and H.
  Wita{\l}a, Phys. Rev. C {\bf 41},  2538  (1990).

\bibitem{kievsky:96a}
A. Kievsky, S. Rosati, W. Tornow, and M. Viviani, Nucl.~Phys. {\bf A607},  402
  (1996).

\bibitem{kievsky:01a}
A. Kievsky, M. Viviani, and S. Rosati, Phys.~Rev.~C {\bf 64},  024002  (2001).

\bibitem{chen:01a}
C.~R. Chen, J.~L. Friar, and G.~L. Payne, Few-Body Syst. {\bf 31},  13  (2001).

\bibitem{deltuva:05a}
A. Deltuva, A.~C. Fonseca, and P.~U. Sauer, Phys.~Rev.~C {\bf 71},  054005
  (2005).

\bibitem{deltuva:05d}
A. Deltuva, A.~C. Fonseca, and P.~U. Sauer, Phys.~Rev.~C {\bf 72},  054004
  (2005).

\bibitem{nollett:07a}
K.~M. Nollett, S.~C. Pieper, R.~B. Wiringa, J. Carlson, and G.~M. Hale, Phys.
  Rev. Lett. {\bf 99},  022502  (2007).

\bibitem{quaglioni:08a}
S. Quaglioni and P. Navratil, Phys. Rev. Lett. {\bf 101},  092501  (2008).

\bibitem{fonseca:87}
A.~C. Fonseca,  in {\em Lecture Notes in Physics 273} (Springer, Heidelberg,
  1987), p. 161.

\bibitem{viviani:01a}
M. Viviani, A. Kievsky, S. Rosati, E.~A. George, and L.~D. Knutson, Phys. Rev.
  Lett. {\bf 86},  3739  (2001).

\bibitem{kievsky:08a}
A. Kievsky, S. Rosati, M. Viviani, L.~E. Marcucci, and L. Girlanda, J. Phys. G
  {\bf 35},  063101  (2008).

\bibitem{viviani:10a}
M. Viviani, R. Schiavilla, L. Girlanda, A. Kievsky, and L.~E. Marcucci, Phys.
  Rev. C {\bf 82},  044001  (2010).

\bibitem{viviani:13a}
M. Viviani, L. Girlanda, A. Kievsky, and L.~E. Marcucci, Phys. Rev. Lett. {\bf
  111},  172302  (2013).

\bibitem{lazauskas:phd}
R. Lazauskas, Ph.D. thesis, Universit\'{e} Joseph Fourier, Grenoble, 2003,
  http://tel.ccsd.cnrs.fr/documents/archives0/00/00/41/78/.

\bibitem{lazauskas:04a}
R. Lazauskas and J. Carbonell, Phys. Rev. C {\bf 70},  044002  (2004).

\bibitem{lazauskas:09a}
R. Lazauskas, Phys. Rev. C {\bf 79},  054007  (2009).

\bibitem{lazauskas:12a}
R. Lazauskas, Phys. Rev. C {\bf 86},  044002  (2012).

\bibitem{deltuva:07a}
A. Deltuva and A.~C. Fonseca, Phys.~Rev.~C {\bf 75},  014005  (2007).

\bibitem{deltuva:07b}
A. Deltuva and A.~C. Fonseca, Phys.~Rev.~Lett. {\bf 98},  162502  (2007).

\bibitem{deltuva:12c}
A. Deltuva and A.~C. Fonseca, Phys.~Rev.~C {\bf 86},  011001(R)  (2012).

\bibitem{deltuva:13c}
A. Deltuva and A.~C. Fonseca, Phys.~Rev.~C {\bf 87},  054002  (2013).

\bibitem{deltuva:07c}
A. Deltuva and A.~C. Fonseca, Phys.~Rev.~C {\bf 76},  021001(R)  (2007).

\bibitem{deltuva:14a}
A. Deltuva and A.~C. Fonseca, Phys.~Rev.~Lett. {\bf 113},  102502  (2014).

\bibitem{deltuva:08a}
A. Deltuva, A.~C. Fonseca, and P.~U. Sauer, Phys.~Lett.~B {\bf 660},  471
  (2008).

\bibitem{deltuva:14b}
A. Deltuva and A.~C. Fonseca, Phys.~Rev.~C. {\bf 90},  044002  (2014).

\bibitem{deltuva:10c}
A. Deltuva, Phys.~Rev.~A {\bf 82},  040701(R)  (2010).

\bibitem{taylor:74a}
J.~R. Taylor, Nuovo Cimento B {\bf 23},  313  (1974); M.~D. Semon and J.~R.
  Taylor, Nuovo Cimento A {\bf 26}, 48 (1975).

\bibitem{alt:80a}
E.~O. Alt and W. Sandhas, Phys.~Rev.~C {\bf 21},  1733  (1980).

\bibitem{kankowsky:76}
R. Kankowsky, J.~C. Fritz, K. Kilian, A. Neufert, and D. Fick, Nucl. Phys. {\bf
  A263},  29  (1976).

\bibitem{brolley:60}
J.~E. Brolley, T.~M. Putnam, L. Rosen, and L. Stewart, Phys. Rev. {\bf 117},
  1307  (1960).

\bibitem{PhysRevC.4.52}
J.~L. Detch, R.~L. Hutson, N. Jarmie, and J.~H. Jett, Phys. Rev. C {\bf 4},  52
   (1971).

\bibitem{pt19-57}
R. Darves-Blanc, N. Sen, J. Arvieux, A. Fiore, J. Gondrand, and G. Perrin,
  Lett. Nuovo Cimento {\bf 4},  16  (1972).

\bibitem{doleschall:04a}
P. Doleschall, Phys.~Rev.~C {\bf 69},  054001  (2004).

\bibitem{machleidt:01a}
R. Machleidt, Phys.~Rev.~C {\bf 63},  024001  (2001).

\bibitem{deltuva:03c}
A. Deltuva, R. Machleidt, and P.~U. Sauer, Phys.~Rev.~C {\bf 68},  024005
  (2003).

\bibitem{hardekopf:72a}
R. Hardekopf, P. Lisowski, T. Rhea, R. Walter, and T. Clegg, Nucl. Phys. A {\bf
  191},  481   (1972).

\bibitem{wilson:61}
W.~E. Wilson, R.~L. Walter, and D.~B. Fossan, Nucl. Phys. {\bf 27},  421
  (1961).

\bibitem{drosg:78}
M. Drosg, Nucl. Sci. Eng. {\bf 67},  190  (1978); in EXFOR Database (NNDC,
  Brookhaven, 1978).

\bibitem{PhysRevC.16.15}
N. Jarmie and J.~H. Jett, Phys. Rev. C {\bf 16},  15  (1977).

\bibitem{allas:74a}
R.~G. Allas, L.~A. Beach, R.~O. Bondelid, E.~M. Diener, E.~L. Petersen, J.~M.
  Lambert, P.~A. Treado, and I. Slaus, Phys. Rev. C {\bf 9},  787  (1974).

\bibitem{jarmer:74}
J.~J. Jarmer, R.~C. Haight, J.~E. Simmons, J.~C. Martin, and T.~R. Donoghue,
  Phys. Rev. C {\bf 9},  1292  (1974).

\bibitem{PhysRevC.10.494}
J.~J. Jarmer, R.~C. Haight, J.~C. Martin, and J.~E. Simmons, Phys. Rev. C {\bf
  10},  494  (1974).

\bibitem{PhysRevC.5.1826}
R.~C. Haight, J.~E. Simmons, and T.~R. Donoghue, Phys. Rev. C {\bf 5},  1826
  (1972).

\bibitem{deltuva:15a}
A. Deltuva and A.~C. Fonseca, Phys. Lett. B, submitted.

\end{thebibliography}

\end{document}